# Distributed Detection in Sensor Networks with Limited Range Multi-Modal Sensors [*]

Erhan Baki Ermis, Venkatesh Saligrama
Department of Electrical and Computer Engineering
Boston University, Boston, MA 02215
Email: {ermis, srv}@bu.edu

## Abstract

We consider a multi-object detection problem over a sensor network (SNET) with limited range multi-modal sensors. Limited range sensing environment arises in a sensing field prone to signal attenuation and path losses. The general problem complements the widely considered decentralized detection problem where all sensors observe the same object. In this paper we develop a distributed detection approach based on recent development of the false discovery rate (FDR) and the associated BH test procedure. The BH procedure is based on rank ordering of scalar test statistics. We first develop scalar test statistics for multidimensional data to handle multi-modal sensor observations and establish its optimality in terms of the BH procedure. We then propose a distributed algorithm in the ideal case of infinite attenuation for identification of sensors that are in the immediate vicinity of an object. We demonstrate communication message scalability to large SNETs by showing that the upper bound on the communication message complexity scales linearly with the number of sensors that are in the vicinity of objects and is independent of the total number of sensors in the SNET. This brings forth an important principle for evaluating the performance of an SNET, namely, the need for scalability of communications and performance with respect to the number of objects or events in an SNET irrespective of the network size. We then account for finite attenuation by modeling sensor observations as corrupted by uncertain interference arising from distant objects and developing robust extensions to our idealized distributed scheme. The robustness properties ensure that both the error performance and communication message complexity degrade gracefully with interference.

## 1 Introduction

The design and deployment of sensor networks (SNET) for distributed decision making pose fundamental challenges due to energy constraints and environmental uncertainties. While power and energy constraints limit collaboration among sensors nodes, some form of collaboration is necessary to overcome uncertainty and meet reliability requirements of the decision making process.

In this paper we focus on the problem of distributed detection of localized events, sources or abnormalities (from here on objects), observed simultaneously over different sections of a large sensor network. Such problems arise naturally in many settings such as environmental monitoring, species distribution and taxonomy, and wide area surveillance [17, 20]. The common thread in all of these applications is that the objects are not observed by all the sensors in the SNET. Rather, each object is in the field-of-view of only a small subset of the sensors in the SNET. We consider all such problems to be local information problems, and seek to devise a distributed detection strategy that satisfies certain false alarm and communication cost constraints.

It is worth contrasting local information problem with its global counterpart. In a global information problem a single object is observed across the entire network (see Figure 1 for an illustration of local and global information problems). This type of problem has been extensively studied in the literature in the context of decentralized/distributed

[*]This research was supported by ONR Young Investigator Award N00014-02-100362, Presidential Early Career Award (PECASE), and NSF Award CCF-0430993

detection theory [6, 21, 24, 26, 28, 30]. In the centralized version of the problem one seeks asymptotically the optimal exponent at which the error probability goes to zero as a function of the observations [23, 27]. The decentralized version involves a similar problem with quantized observations [28, 30]. Motivated by network topologies researchers have also investigated several architectures ranging from fusion centric to ad-hoc consensus based approaches [3, 6, 18, 21, 22, 24, 26, 28–30]. Local information problems and corresponding decentralized algorithms have only recently begun to be addressed in the SNET setting [12–15]. A fundamental difference between local and global information problems appears even in the centralized scenario. In the local information case, since each object is in the field-of-view of at most a constant number of sensors the error probability cannot be made to go to zero. Furthermore, since there are a multiple locations each location has to be simultaneously tested for presence or absence of objects. In these cases neither the total number of objects in the sensor field nor the likelihood of finding an object in a specific location is known a priori. It turns out that in these cases the error probability is dominated by the multiple tests (one for each location) and this issue is referred to as multiple comparisons testing in the statistical literature [4, 19]. A fundamental difference is in what performances are typically characterized. While, for global information problems, the asymptotics of the error probability for a single object with increasing number of sensors (and quantized observations) is usually derived, for local information problems, the scalability of error rates and communication costs with increasing number of objects is characterized. An important aspect of our work is to show that both of these quantities, namely the error rates and communication costs, scale with the number of sensors that are in the immediate vicinity of objects rather than the size of the SNET.

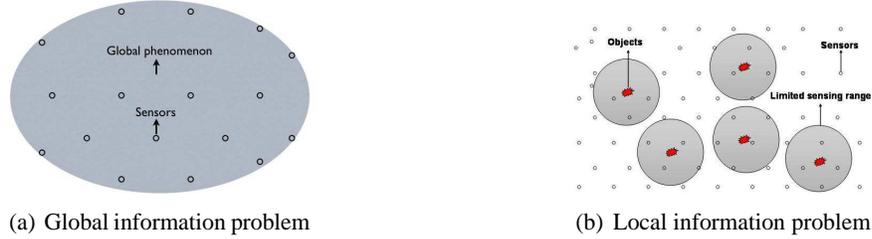

(a) Global information problem    (b) Local information problem

Figure 1: In global information problems the sensors observe a single global phenomenon, which leads to a binary hypotheses testing problem with multiple observations. In local information problems only a subset of the sensors observe a number of phenomena, which leads to each sensor having its own set of hypotheses. This leads to a multiple hypotheses testing problem.

We present a distributed detection scheme for local information problems based on the concept of false discovery rate and the associated BH procedure [4]. The BH procedure relies on rank-ordering of test statistics. In several SNET scenarios multi-modal sensors are employed, which generate multi-dimensional sensor observations, where rank ordering is unclear. We also consider a sensing field with signal attenuation and path losses, which essentially imposes an effective sensing range for the sensors.

To the best of our knowledge multi-dimensional settings in the context of FDR have not been subjected to significant attention since it is generally difficult to rank-order the observations. Recent statistical work in [7] proposes a coordinate-by-coordinate ordering but this generally leads to sub-optimal error performance. To account for multi-dimensional observations we devise a transformation that maps multidimensional observations to scalar test statistics, which turns out to have optimal error performance. These scalar statistics then forms a basis for a distributed detection scheme. We show that the communication cost of the scheme scales linearly with the number of sensors that observe an object, and not the number of sensors that are in the SNET. Furthermore, the proposed scheme guarantees detection performance of centralized procedures. Next, we account for signal attenuation and path losses by modeling sensor observations as corrupted by uncertain interference resulting from unknown objects that are not in the immediate vicinity of the sensor. The interference can be modeled as a perturbation to the nominal observed distributions and we establish robustness of our test statistic to such perturbations.

The organization of the paper is as follows: in Section 2 we discuss the connection of distributed detection of localized phenomena to the multiple hypotheses testing problems considered in the statistical literature. We discuss possible performance criteria in detail and present the reasoning behind our choice. We also discuss the main contributions of this work in that section. In Section 3 we discuss the setup of our problem and describe ideal and non-ideal sensing models. In Section 4 we propose a test statistic formulation, and discuss its important properties. In Section 5 we present the distributed detection algorithm, and examine its scaling properties. We also show here that the distributed algorithm is equivalent to its centralized counterpart with high probability. We then show, in Section 6, certain robustness properties of the test statistics to uncertainties in the distribution of observations. We also show that our choice of performance criterion scales gracefully with the perturbation of the distribution of observations. These results allow us to address cases where we do not have the exact distributions. In Section 7 we present simulations and show that the chosen method is able to meet the Bayes Oracle error performance. In this section we also present the scaling of communication costs and discuss some interesting results. We finally present our concluding remarks in Section 8.

## 2 Discussion of Performance Metrics and Contributions

In this section we will propose different criteria and present empirical evidence for adopting the Benjamini-Hochberg (BH) procedure, which is associated with the false discovery rate(FDR) criterion, as a basis for local information problems. Local information problems invariably turn out to be multi-comparison test problems. There is currently no consensus around a universally applicable performance metric for these problems. In the literature, location-by-location Neyman Pearson (NP) tests, Family-Wide-Error (FWER, also known as Bonferroni criteria) tests, average error probability and false discovery rate have all been proposed. Rather than discuss merits of the different criteria we describe their performance in terms of average errors for our context, wherein both the object density as well as observed distributions may only be partially known. The NP tests and FWER criterion are non-adaptive decision rules (i.e. threshold rules which do not depend on observed realization). Generally these methods result in poor performance in terms of the number of false alarms and missed detections. It turns out that the BH procedure, in contrast, is an adaptive rule which adapts to the observed realization and generally results in good error performance.

To be concrete, consider a set of $m$ sensors, $\mathcal{S}$. Associated with each sensor $s \in \mathcal{S}$, there is a null or alternative hypothesis $H^s \in \{H_0, H_1\}$ corresponding to whether or not the sensor observes an object of interest. Sensor $s$ generates an observation $X_s \in \mathbb{R}$ independently (of other sensors) with probability density $g_{0s}$ if $H^s = H_0$; and $g_{1s}$ if $H^s = H_1$. The general problem involves situations where the actual number of objects are unknown and due to path losses and multi-path effects the distributions $g_{0s}$, $g_{1s}$ are only partially known. To analyze different strategies we denote $u(x_1, x_2, \ldots, x_m)$ to be any decision rule that selects a set of sensors $\mathcal{S}_1 = \{s_1, s_2, \ldots, s_R\}$ and assigns to them the alternative hypothesis, i.e., $H^s = H_1$, $s \in \mathcal{S}_1$ and $H_0$ otherwise. The outcome of a decision rule can be summarized in the following table. Here $R$ is the total number of sensors identified with objects. $V$ is the number of sensors falsely placed into $\mathcal{S}_1$, i.e., number of false alarms. Obviously, we desire both $V$ and $T$ to be small, and seek a decision rule that makes this possible.

|  | Declared $H_0$ | Declared $H_1$ | Total |
|---|---|---|---|
| True $H_0$ | $U$ | $V$ | $m_0$ |
| True $H_1$ | $T$ | $Z$ | $m - m_0$ |
| Total | $m - R$ | $R$ | $m$ |

### 2.1 Controlling False Alarm Probability: Non-Adaptive Strategies

First consider the situation when the distributions $g_{0s}$, $g_{1s}$ are known but $m_1$ is unknown and arbitrary. In this case we can consider several possibilities. (a) Neyman-Pearson Test for each location: Maximize local detection power $P_D^l$ subject to local false alarm probability constraint, $P_F^l \leq \gamma$ for each location. The optimal decision rule for this situation is the well-known likelihood ratio test [27]. Although this rule is locally optimal, it is not guaranteed to provide good overall performance and is commonly referred to as uncorrected testing. Indeed the false alarms scale with the total number of sensors, i.e., $E(V) \approx \mathcal{O}(m)$. (b) Bonferroni procedure [4] overcomes this issue by imposing a highly restrictive false alarm probability constraint on each sensor, $\gamma' = \gamma/m$. This strategy guarantees control of global false alarm probability (i.e. $\Pr(V \geq 1)$[1]) at level $\gamma$ (follows from union bound), however this in turn leads to poor detection performance, i.e, a large number of misses are incurred.

This leads to the fundamental question of whether there exist other local or global decision rules, $u(\cdot)$, that can control both the false alarm and miss probability $\Pr(V \geq 1)$, $\Pr(T \geq 1)$ (or a close relaxation such as the probability of $k$ false alarms and misses for some constant $k$ independent of $m$).

The optimal decision rule for maximizing the worst-case global detection power $P_D = 1 - \Pr\{T \geq 1\}$ subject to a global false alarm constraint $P_F = \Pr\{V \geq 1\}$ is generally intractable. It turns out that the worst-case false alarm and miss probability can be bounded from below by an entropic term which is a function only of the local SNR.

**Theorem 2.1** *Let S be a set of $m$ hypotheses tests, $H^s \in \{H_0, H_1\}$ the hypothesis for test $s \in \mathcal{S}$, and $X_s$ the observation for the test $s \in \mathcal{S}$. Suppose, $u(X_1, X_2, \ldots, X_m)$ is any strategy that maps the observations to hypothesis decisions. Then,*

$$\gamma_w = \min_u \max_{H^s \in \{H_0, H_1\}} (Pr\{V \geq 1 \mid \{H^s : s \in \mathcal{S}\}\} + Pr\{T \geq 1 \mid \{H^s : s \in \mathcal{S}\}\}) \geq \Phi(H^s \mid X_s) - \frac{1}{m}$$

*where $\Phi(\cdot \mid \cdot)$ is the conditional entropy computed with a Bernoulli prior with probability $1/2$ over $H_1$s and $H_0$'s over the $m$ tests. It follows that there exists no decision strategy for which both false alarm and miss probability can simultaneously be smaller than $\Phi(H^s \mid X_s)/2$.*

Proof: See Appendix. ∎

**Remark:** $H^s$ is a binary random variable and so its entropy (or conditional entropy) is always smaller than one. Nevertheless, depending on measurement noise at each sensor, $\Phi(H^s \mid X_s)$ could be arbitrarily close to one.

**Remark:** We can generalize this result to lower bound probability of $k$ false alarms and misses as well using generalized Fano bounds we developed in [2]. Based on those results it follows that the probability does not improve unless we let either $V$ or $T$ grow with $m$.

The above discussion brings to light the fact that non-adaptive decision rules lead to poor performance.

## 2.2 Adaptive Strategies

To establish performance of adaptive strategies, i.e., strategies that adapt to the specific realization, we need lower bounds on error performance. We do this by means of a Bayes Oracle where the distributions $g_{0s}$, $g_{1s}$ as well as the likelihood probability of finding an object in the vicinity of a sensor is known (alternatively, we can consider situations where the total number of objects are known). Define, the average ratio, $m_1/m$ as the object density and the complement, namely, average of $m_0/m$ as the sparsity level. Under this scenario it is easy to see that a thresholding decision at each sensor is optimal, and the optimal threshold is a function of the object density and the distributions under each hypothesis. Furthermore, the error performance of the Bayes Oracle is a lower bound on the achievable error probability.

The question therefore arises as to whether there exists a procedure that achieves Bayes Oracle bound for local information problems, and does not depend on the knowledge of the object density and precise knowledge of distribution under presence of an object. This is particularly relevant since path losses and attenuation are not precisely known. Motivated by these issues Benjamini & Hochberg [4] formulate the false discovery rate (FDR) criterion and provide a distribution invariant algorithm, the Benjamini-Hochberg (BH) procedure, that controls FDR. An interesting result presented in [1] shows that controlling FDR can asymptotically result in asymptotic minimax optimality of the error probability.The FDR [4] framework seeks to control the worst case expected ratio of $V/R$, i.e.

$$FDR = \max_{\{H^s \in \{H_0, H_1\}\}_{s \in \mathcal{S}}} E\{V/R \mid \{H^s\}_{s \in \mathcal{S}}\}$$

---

[1]Strictly speaking, we should write $\max_{H^s \in \{H_0, H_1\}} \Pr\{V \geq 1 \mid \{H^s\}_{s \in \mathcal{S}}\}$ to denote that we are looking at worst-case probability. However we avoid this cumbersome notation whenever clear from context.

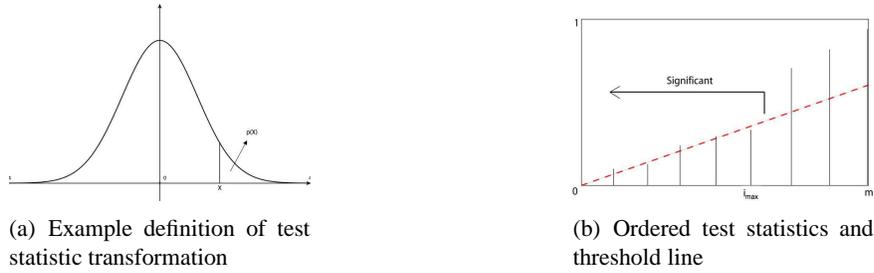

(a) Example definition of test statistic transformation

(b) Ordered test statistics and threshold line

Figure 2: BH procedure

where $\mathcal{S}$ is the set of sensors and $H^s$ is the hypothesis at sensor $s$. For simplicity of notation, from here on we will write $FDR = E\{V/R\}$ and $\Pr\{\cdot \mid \{H^s\}_{s\in\mathcal{S}}\} = \Pr\{\cdot\}$ whenever it is clear from the context. It is easy to show that, $FDR \leq \Pr\{V \geq 1\}$, which implies that FDR is a relaxation of global false alarm probability.

FDR can be controlled using the so called BH procedure, which we briefly explain here. As depicted in Figure 2 the test statistics are computed from the observations. The test statistic of an observation is obtained through any (*non-unique*) transformation that generates a uniform distribution, $U[0,1]$, under null hypotheses. The test statistics are then rank ordered and a desired FDR threshold, $\gamma$, is chosen. Let $y_{(i)}$ be the $i^{th}$ smallest test statistic. The largest index $i_{max}$, such that $y_{(i)} \leq \frac{i}{m}\gamma$ is chosen as the decision point, and the test statistics whose rank indices are smaller than $i_{\max}$ are labeled significant, i.e., mapped to alternative hypotheses. The BH procedure ensures that $FDR \leq \gamma$ for a desired threshold $\gamma$, regardless of how the observations under $H_1$ are generated.

Thus the BH procedure [4, 5] is an adaptive thresholding procedure and the final stopping point is itself a random variable [16] and depends on the specific realization. Nevertheless, it can be shown that the BH procedure [4] is a distribution invariant algorithm (i.e., regardless of $g_{1s}$) controls FDR below $\gamma$.

**Theorem 2.2** *For independent test statistics under null hypothesis, and for any configuration of alternative hypotheses, the BH procedure controls the FDR at level $\gamma m_0/m$, where $m_0$ is the number of true null hypothesis and $m$ is the number of observations.*

For our purposes error performance of the BH procedure is of relevance. In [16] it is shown that the BH procedure achieves the Bayes Oracle performance for reasonable signal-to-noise ratio and low-levels of target density even though the distribution under $H_1$ (they impose weak conditions on $g_{1s}$) as well as actual number of objects maybe unknown. A related result in [30] shows that adaptive procedures such as the BH procedure outperform fixed threshold procedures. As seen in Figure 3, when we control the FDR criterion, the error rate closely tracks the error performance of the Bayes Oracle risk policy. In conclusion, the above exercise shows that adaptive procedures adapt their threshold to object density in contrast to non-adaptive procedures. In addition they have inherent robustness properties that can be useful in our SNET setting. With this justification we adopt the FDR framework for our problem.

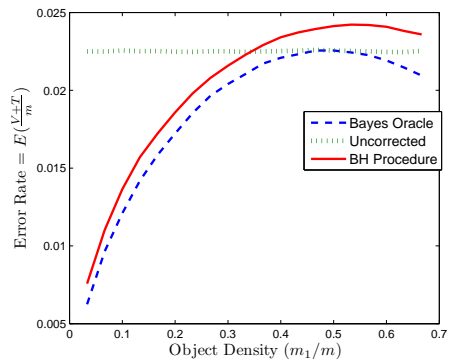

Figure 3: Monte Carlo error rate comparison of Uncorrected Testing, Bayes Oracle, and the BH procedure for varying object density. 150 samples with distributions $N(0,1)$ under $H_0$ and $N(0,4)$ under $H_1$ were used with varying target density.

## 2.3 Contributions

First, although BH procedure ensures FDR control irrespective of the observed distribution under presence of object, the detection performance can vary widely for different distributions. In particular, the reference distribution corresponding to the null hypotheses can be transformed in several ways to obtain test statistics without changing the FDR control. This leaves room for optimizing detection performance. The problem is particularly acute in multi-dimensional settings since it is difficult to rank-order the observations. BH procedure works on scalar test statistics, where there is a clear ordering. In [7] a coordinate-by-coordinate ordering is presented but this generally is sub-optimal. Our first contribution is to present a transformation that maps multi-dimensional observations to scalar test statistics and enables rank-ordering of the observations and application of BH procedure. We show that our test statistics are optimal in the sense that it leads to maximal detection power for a given level of FDR control.

A second contribution of our work is to show that our transformations and procedures are robust to perturbations of the distribution of observations. This is particularly important in the SNET setting due to path-loss effects. Indeed, due to complex nature of attenuations and randomness in the environment, signals from objects far away can interfere with signals from objects in the immediate vicinity. However, the interfering signal is unknown and this motivates development of robust techniques.

Our third contribution is in developing a distributed, communication efficient BH procedure for multi-object detection for SNETs. Our results indicate that *corresponding to an FDR threshold, the communication message complexity grows in proportion to the actual number of sensors observing objects (significant sensors) while achieving the same centralized performance*. Namely, the communication costs scale with event density for a pre-specified error performance and is independent of the network size.

## 3 Setup

We consider a non-Bayesian setting where an unknown number of objects are distributed on a sensor field of $m$ sensors. We consider a scheme in which the objects generate a signal field over the sensor network and the sensors sample the field at their locations. In this scheme the significant hypothesis ($H_1$) for a sensor is the event that the sensor is within a radius $d_0$ of an object, and the null hypothesis ($H_0$) is the event that the sensor is outside a radius $d_0$ from all objects. We assume a sparse distribution of objects, i.e., at most a single object is allowed to be present in the immediate vicinity of any sensor (we will comment on how to generalize the analysis to handle multiple objects in the immediate vicinity in the following section). Note however that each object can be in the immediate vicinity of multiple sensors.

We call the radius $d_0$ as the effective sensing range of a sensor. This is the radius within which signal energy does not decay significantly. Note that this situation models both active and passive sensing scenarios. In active sensing, sensors transmit a waveform and the return signal undergoes path losses. In passive sensing objects radiate signal patterns, which undergo path losses as well. Therefore, the object being in the vicinity of a sensor or the sensor being in the vicinity of an object are mathematically equivalent. The observations at each sensor are multidimensional to account for multi-modal sensors with different modalities such as magnetic, seismic, and acoustic.

In this work we separate the problem of *what to communicate* from the problem of *how to communicate* by assuming a broadcast model, wherein each sensor, once it decides on what to communicate, broadcasts that information to the entire SNET. The reason for separating these two problems is, given we know what we want network to compute, there are a number of methods that offer an efficient solution and only require communication connectivity [11, 31]. Consequently, communication complexity is the aggregate number of messages broadcast by the sensors. Our objective is a distributed decision rule, which has low communication complexity and good error performance.

### 3.1 Mathematical Modeling of Multi-Modal Sensor Observations

We begin by considering an example of the following sensing model:

$$\mathbf{X}_s = \sum_t \frac{1}{(\frac{d(s,t)}{d_{min}})^\alpha + 1} \boldsymbol{\theta}_t + \boldsymbol{\nu}_s \tag{1}$$

where $\boldsymbol{\theta}_t$ is the multidimensional signal (possibly random with known distribution) of object $t$, $d(s,t)$ is the distance between sensor $s$ and an object $t$. The minimum distance $d_{min}$ is the distance below which the path loss model does not hold and the signal saturates. The model above (with the $d_{\min}$ and one in the denominator) is a simplified model to account for both near field and far-field effects and ensures that the received signal power is not larger than radiated power, whenever the object is in the close vicinity of a sensor. The parameter, $\alpha$, is the power decay exponent for the path loss, $\boldsymbol{\nu}_s$ is the multidimensional noise variable of known distribution.

Note that each sensor can consist of multiple modalities such as Electromagnetic (EM), Acoustic(AC) and Seismic (SE) etc. Thus, with $'$ denoting transpose, the parameter (possibly random with known distribution) $\theta_t$ above can be decomposed as

$$\boldsymbol{\theta}_t = (\theta_t^{EM}, \theta_t^{AC}, \theta_t^{SE})'$$

Note that $d_{min}$ can be also be different for different modalities, however for simplicity of notation we have assumed that it is the same for all modalities. With this observation model, we have the hypotheses as follows. Note that observations at each sensor are conditionally independent when conditioned on the underlying hypothesis.

$$\begin{aligned} H^s &= H_0: \ d(s,t) > d_0 \ \text{ for all objects } t \\ H^s &= H_1: \ d(s,t) \leq d_0 \ \text{ for an object } t \end{aligned}$$

The distance $d_0$ is typically chosen to be the distance where the signal power relative to noise power is sufficiently large. Therefore, in general $d_0$ is close to $d_{\min}$ for large attenuation coefficients ($\alpha$).

For simplicity of exposition we assume in this paper that only one object can be present within the distance $d_0$. This is usually satisfied when we have a sparse objects distributed in the sensing field. However, we briefly discuss how these techniques can be generalized to handle multiple objects within $d_0$. We point out that multiple objects can be incorporated by using an extended hypothesis space along the lines of [8]. There are two cases here to consider: (a) Multiple objects lead to sufficiently different signal patterns; (b) Multiple objects do not lead to sufficiently different discrimination. In the first case the hypothesis space can be expanded to account for multiple objects. The main idea is to have multiple null and significant hypotheses for each sensor, $s$. The kth null hypothesis, $H_{k0}$ at sensor $s$ corresponds to the hypothesis that there are less than $k$ objects in the vicinity of the sensors, while the kth significant hypothesis, $H_{k1}$, corresponds to the hypothesis that there are exactly $k$ objects in the vicinity. For observations distributed as exponential random variables, the generalized maximum likelihood test statistics are independent conditioned on the null hypothesis. This fact is sufficient to apply Theorem 2.2 and quantify performance of BH procedure. Thus this idea can be integrated with the distributed sensors to form a expanded hypothesis set, which meets the conditions required of BH procedure. For the second case when multiple objects do not result in sufficiently different signal patterns the robustness techniques developed in the paper apply.

In summary, hypotheses are associated with each sensor. Hypothesis $H_1$ at sensor $s$ corresponds to existence of an object within a radius $d_0$ of a sensor, while hypothesis, $H_0$ corresponds to no object within radius $d_0$. An important point to note here is that each object can be in vicinity of multiple sensors and so the hypothesis $H_1$ at multiple sensors can result from the same object.

**Ideal Sensing Model(Case of Infinite Attenuation):** Note that as the attenuation coefficient $\alpha$ gets larger, the distribution of observations takes a nominal form, where within a radius $d_{min}$ of an object $t$, the received signal has a mean $\boldsymbol{\theta}_t$, (if $\boldsymbol{\theta}_t$ is non-random parameter) and outside this radius the received signal has negligible mean. Thus sensors within the minimum radius, $d_{\min}$ of the object receive the full signal power, while sensors outside this radius receive negligible power. This leads to the following simplified model:

$$\begin{aligned} H^s &= H_0: \ \mathbf{X}_s = \boldsymbol{\nu}_s \\ H^s &= H_1: \ \mathbf{X}_s = \boldsymbol{\theta}_t + \boldsymbol{\nu}_s \end{aligned}$$

Throughout the paper we refer to the first model as the non-ideal sensing model, and the limiting case as the Ideal Sensing Model. Figure 4 illustrates these models for a hypothetical mode.

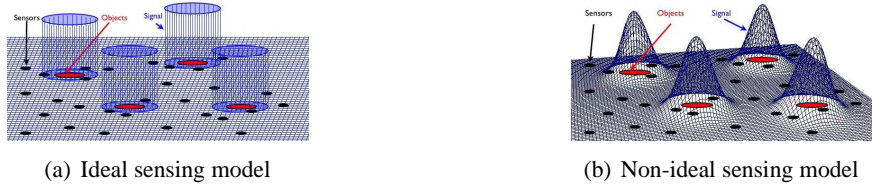

(a) Ideal sensing model  (b) Non-ideal sensing model

Figure 4: Ideal and non-ideal sensing models: In ideal sensing model the objects have constant signal over a confined region, whereas in the non-ideal sensing model the signal decays and is not confined.

These examples can be abstracted to a more general formulation, where the noise is no longer modeled as additive and $\boldsymbol{\theta}_t$ can belong to an arbitrary distribution. The cumulative probability distribution (resp. density) function of the observation vector $\mathbf{X}_s$ at sensor $s$ under each hypothesis $H^s = H_i$, $i = 0, 1$ is denoted by $G_{is}(\cdot)$ (resp. $g_{is}(\cdot)$), where $H^s$ denotes the hypothesis at sensor $s$. In the non-ideal case we have two composite family of continuous distributions, $g_{is}(\cdot) \in \mathcal{G}_i$, $i = 1, 2$. Observations at each sensor is conditionally independent when conditioned on the underlying hypothesis. The observations are denoted by a vector $\mathbf{X}_s = (X_s^1, X_s^2, \ldots, X_s^d)$, $s \in \mathcal{S}$, where $d$ is the number of dimensions, $X_s^j$ represents the $j^{th}$ dimension of the measurement taken by sensor $s \in \mathcal{S}$, and $\mathcal{S}$ represents the set of sensors that form the SNET. The realization of observation vector $\mathbf{X}_s$ is denoted by $\mathbf{x}_s = (x_s^1, x_s^2, \ldots, x_s^d)$, $s \in \mathcal{S}$.

$$H^s = H_0 : \mathbf{X}_s \sim g_{0s} \in \mathcal{G}_0$$
$$H^s = H_1 : \mathbf{X}_s \sim g_{1s} \in \mathcal{G}_1$$

We let $\mathcal{S}_0 = \{s \in \mathcal{S} : H^s = H_0\}$ with cardinality $m_0$ and $\mathcal{S}_1 = \{s \in \mathcal{S} : H^s = H_1\}$ with cardinality $m_1$. Here both $m_0$ and $m_1$ are unknown and the object locations are assumed to be arbitrary, i.e. not necessarily uniformly distributed.

Note that the class of distributions $\mathcal{G}_{0s}$ and $\mathcal{G}_{1s}$ are singletons in the ideal model, regardless of the dimensionality of the observed signal, if we assume that no two objects are within the radius $d_{\min}$ of a single sensor. As we discussed earlier, this is usually true for sparse set of objects in a sensing field. Our approach is to develop results first for the ideal sensing model, where the families of distributions under the two hypothesis are characterized by singletons. We deal with the more general case of $\alpha < \infty$ from a robustness perspective, i.e., as a perturbation of the ideal sensing model, in the upcoming parts of the paper.

## 4 Proposed Test Statistics

In this section we describe the proposed statistic and establish some important properties. We show that these test statistics can be used to perform detection through BH procedure, and allow for control of FDR at desired levels. With our definition, the CDF of test statistics under significant hypotheses becomes a concave function. Based on this concavity property, we can devise a scalable distributed procedure that achieves the detection power of its centralized counterpart. In the remainder of this section we propose a definition of test statistics. The test statistics transform multi-dimensional observations to scalar statistics and are based on volumes of level sets of the likelihood ratio function (more precisely the Radon-Nikodym derivative). These test statistics result in:

(a) The test statistics under null hypotheses are uniformly distributed in $[0, 1]$,

(b) The test statistics under significant hypotheses are "maximally" clustered around zero. Consequently, thresholds near zero lead to detections with relatively few false alarms.

To this end, let $\mu_{0s}$ and $\mu_{1s}$ be the measures associated with the distributions $G_{0s}$ and $G_{1s}$ respectively. We assume throughout this work that $\mu_{1s}$ is absolutely continuous with respect to $\mu_{0s}$, denoted $\mu_{1s} << \mu_{0s}$. Let $\phi_s = d\mu_{1s}/d\mu_{0s}$ be the Radon-Nikodym derivative, i.e. the likelihood ratio function. Define the following transformation of the random variable $\mathbf{X}_s$ from $n$ dimensional space onto the one dimensional space:

$$Y_s = \chi_s(\mathbf{X}_s) = \mu_{0s}\{\mathbf{x} : \phi_s(\mathbf{x}) > \phi_s(\mathbf{X}_s)\} = \mu_{0s}\{\mathbf{x} : \phi_s(\mathbf{x}) \geq \phi_s(\mathbf{X}_s)\} \qquad (2)$$

where we assume that $\phi_s$ is nowhere constant[2]. The nowhere constant assumption holds for example when the involved distributions are Gaussians with different means or different variances. In this definition, the set $\{\mathbf{x} : \phi_s(\mathbf{x}) > \phi_s(\mathbf{X}_s)\}$ is the most powerful decision region such that the probability of false alarm is less than some $\gamma_0$; i.e., it is the solution to $\arg\max_A \mu_{1s}(A)$ subject to $\mu_{0s}(A) \leq \gamma_0$ for some $\gamma_0 \in (0,1)$. Similarly, let $\mu_{1s}\{\mathbf{x} : \phi_s(\mathbf{x}) > \phi_s(\mathbf{X}_s)\} = \gamma_1$ for some $\gamma_1$. Then the set $\{\mathbf{x} : \phi_s(\mathbf{x}) > \phi_s(\mathbf{X}_s)\}$ is also the solution to $\arg\min_A \mu_{0s}(A)$ subject to $\mu_{1s}(A) \geq \gamma_1$; i.e., it is the volume of the so called minimum volume sets at level $\gamma_1$. All of these results follow from the fact that the Radon-Nikodym derivative is precisely the likelihood ratio function.

We now give an example to depict graphically the impact of transformation on a one dimensional distribution. Let us assume that we are given two distributions, whose densities are $g_{0s}$ and $g_{1s}$ as depicted in Figure 5 (a). We first calculate the Radon-Nikodym derivative $\phi_s$, as depicted in Figure 5 (b). For a given $\mathbf{X}_s$ we can now obtain $\phi_s(\mathbf{X}_s)$. We can next identify the set $\{\mathbf{x} : \phi_s(\mathbf{x}) > \phi_s(\mathbf{X}_s)\}$, and obtain $Y_s = \chi_s(\mathbf{X}_s) = \mu_{0s}\{\mathbf{x} : \phi_s(\mathbf{x}) > \phi_s(\mathbf{X}_s)\}$, as depicted in Figure 5. The same intuition holds for multidimensional observations as well. The problem of obtaining

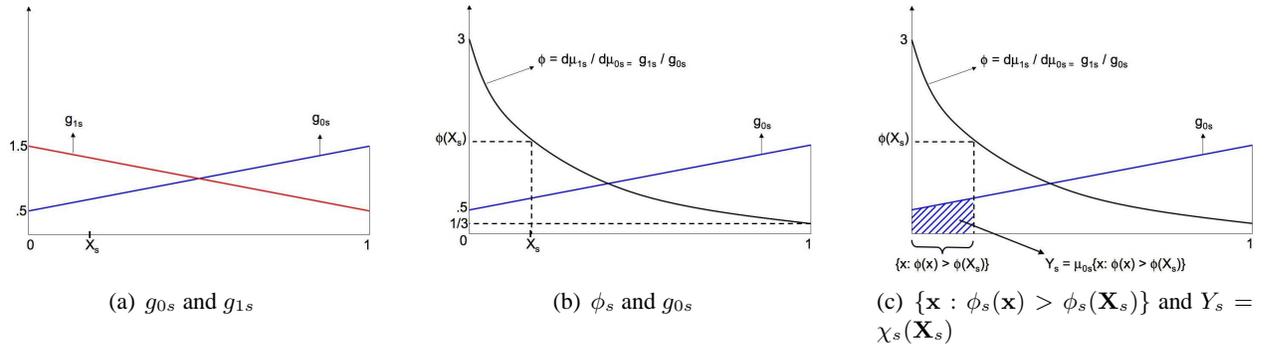

(a) $g_{0s}$ and $g_{1s}$   (b) $\phi_s$ and $g_{0s}$   (c) $\{\mathbf{x} : \phi_s(\mathbf{x}) > \phi_s(\mathbf{X}_s)\}$ and $Y_s = \chi_s(\mathbf{X}_s)$

Figure 5: Depiction of how to obtain the test statistic with a 1-dimensional example.

test statistics for multidimensional random variables has received attention from various researchers. It is worth noting that the test statistics we propose through the transformation are distributed $U[0,1]$ under null hypotheses (which we establish in the following section). Note that transformations that map null distributions to uniform distribution is not unique. For example, in [7], the authors propose to obtain test statistics in a dimension-by-dimension manner, and in [25] a minimum volume set approach is taken. Some of these transformations are compared in Section 7. Our method is relatively simple to implement and guarantees optimal FDR performance in comparison.

## 4.1 FDR Control Using $\chi$

We now establish that using test statistics obtained through $\chi$ as input to BH procedure guarantees FDR control below any desired level. We show this by establishing that under the null hypothesis the test statistics $Y_s = \chi(\mathbf{X}_s)$ are distributed uniformly.

Let $Y_{0s} = \chi_s(\mathbf{X}_s) \sim F_{0s}$ be the random variable when $\mathbf{X}_s \sim G_{0s}$ and $Y_{1s} = \chi_s(\mathbf{X}_s) \sim F_{1s}$ be the random variable when $\mathbf{X}_s \sim G_{1s}$. We begin by establishing that $Y_{0s} \sim U[0,1]$, which implies that we can control FDR at desired levels through BH procedure if we use the $Y_s$ as test statistics.

**Theorem 4.1** *If the derivative $\phi_s = d\mu_{1s}/d\mu_{0s}$ is nowhere constant, then $Y_s = \chi_s(\mathbf{X}_s)$ of Equation 2 is uniformly distributed in $[0,1]$; i.e., $Y_s \sim U[0,1]$.*

Proof: See appendix. ∎

**Corollary 4.2** *BH procedure applied to $Y_s = \chi_s(\mathbf{X}_s) \in [0,1]$, $s = 1, 2, \ldots, m$ of Equation 2 controls FDR.*

---

[2] If $\phi_s$ is constant in some regions, then the transformation can be modified to $Y_s = \chi_s(\mathbf{X}_s) = \mu_{0s}\{\mathbf{x} : \phi_s(\mathbf{x}) > \phi_s(\mathbf{X}_s)\} + \psi$, where $\psi \sim U(0, \beta)$ is a random variable used as a dither with amplitude $\beta_s = \mu_{0s}\{\mathbf{x} : \phi_s(\mathbf{x}) = \phi_s(\mathbf{X}_s)\}$. This dither achieves is analogous to randomized decision rules used in detection theory [27]. The results developed in the paper are valid for this general case but the proofs are more involved.

As we mentioned in Section 2, the main insight behind the proof of Theorem 2.2 lies in the simple fact that the test statistics under null hypotheses are independent and uniformly distributed under null hypothesis. Then, the corollary follows readily from Theorem 4.1.

## 4.2 Optimality Results for Transformation $\chi$

Here we prove that $\chi$ is the optimal transformation in the sense that it maximizes the detection power of BH procedure subject to any FDR constraint $\gamma$. We also establish that under the significant hypothesis, the distribution of $Y_s = \chi(\mathbf{X}_s)$ is concave, which leads to the important result that the optimal decision rule in the space of $Y_s$ is a thresholding rule. This result carries importance from the distributed detection perspective, as will be clear in the upcoming sections.

The following theorem will be necessary to formalize the fact that our proposed transformation of Equation 2 leads to maximal detection power of BH procedure.

**Theorem 4.3** *Let $Z_s = \hat{\chi}_s(\mathbf{X}_s)$ be any test statistic obtained from the observations $\mathbf{X}_s$ such that $Z_s \sim U[0,1]$ under the null hypothesis. If $Y_s = \chi_s(\mathbf{X}_s)$ of Equation 2, then $Pr\{Y_s \leq \gamma_s\} \geq Pr\{Z_s \leq \gamma_s\}$, where the probability measure is $\mu_s = \pi\mu_{1s} + (1-\pi)\mu_{0s}$ for some mixture parameter $\pi$.*

Proof: See appendix. ∎

We next establish that $F_{1s}$, the distribution of $Y_{1s}$, is concave; i.e. the density function $f_{1s}$ is monotone decreasing in $[0, 1]$. This result has strong implications in terms of the detection power and scalability of the distributed algorithm.

**Theorem 4.4** *If the derivative $\phi_s = d\mu_{1s}/d\mu_{0s}$ is nowhere constant, then $F_{1s}$, is concave.*

Proof: See appendix. ∎

We next present an optimality result over a family of testing procedures. Suppose, $u_\gamma(\cdot)$, $\gamma \in [0, 1]$ is a family of testing procedures such that $u_\gamma(\cdot)$ controls the false alarm at level $\gamma$. Let, $\Lambda_\gamma$ be the set of observations, $\mathbf{X}_s$ that are accepted as significant, i.e.,

$$\Lambda_\gamma = \{\mathbf{X}_s \;:\; u_\gamma(\mathbf{X}_s) = H_1\}$$

For each observation, $X_s$ define the mapping,

$$\hat{\chi}(\mathbf{X}_s) = \inf_\gamma\{\gamma : \mathbf{X}_s \in \Lambda_\gamma\} \in [0, 1] \tag{3}$$

It is easy to check that under suitable technical conditions the mapping is a Borel measurable function and induces a uniform measure on $[0, 1]$. We then have the following theorem.

**Corollary 4.5** *Let $R = [0, \gamma]$ be a decision region such that if $Y_s \in R$ we decide $H^s = H_1$, otherwise we decide $H^s = H_0$. If $Y_s$ are obtained through transformation $\chi(\cdot)$, the decision region $R$ maximizes probability of detection subject to a probability of false alarm constraint $\gamma$ over any other family of decision rules $u_\gamma(\cdot)$ defined above.*

Corollary 4.5 follows immediately from Theorems 4.3 and 4.4. Recall that $F_{0s}$ is a uniform distribution in $[0, 1]$, any set (in $[0, 1]$) of Lebesgue measure $\gamma$ has probability of false alarm exactly $\gamma$. Since, according to Theorem 4.4, $F_{1s}$ is concave, its density is monotone decreasing. Therefore among the sets of length $\gamma$, $R = [0, \gamma]$ carries the most mass under $F_{1s}$. Furthermore, as a consequence of Theorem 4.3, among all transformations that generate a uniform $F_{0s}$, $\chi$ maximizes $F_{1s}(\gamma)$ and the corollary follows.

We now state an important corollary regarding the maximal detection power of BH procedure with the proposed test statistics.

**Corollary 4.6** *The BH procedure when applied to $\chi$ of Equation 2 is larger that any other transformation $\hat{\chi}$ for which the null distribution is also $U[0, 1]$.*

Note that in the BH procedure, the test statistics $Y_s$ is compared against a threshold $\gamma_s$, this result follows immediately from Corollary 4.5.

## 4.3 Convexity Properties of Ordered Test Statistics

We present another important implication of Theorem 4.4 next. We show that the expected value of ordered test statistics are samples of a convex function, an important property for designing the scalable distributed detection algorithm. In fact, it is due to this result that we can achieve the centralized performance through a decentralized method.

**Corollary 4.7** *Let $Y_{(1)}, Y_{(2)}, \ldots, Y_{(m)}$ be the rank ordered test statistics such that $Y_{(i)}$ is the $i^{th}$ smallest of $Y_s$, $s = 1, 2, \ldots, m$. $E(Y_{(i)})$, $i = 1, 2, \ldots, m$ are samples of a convex function, i.e., $E(Y_{(i)}) \leq (E(Y_{(i-1)}) + E(Y_{(i+1)}))/2$, $i = 2, 3, \ldots, m-1$ asymptotically, as $m \to \infty$.*

Asymptotically, the $E(Y_{(i)}) = F_s^{-1}(i/(m+1))$ [9] where $F_s = \pi F_{1s} + (1-\pi) F_{0s}$ for any mixing parameter $\pi$. But, according to Theorem 4.4, $F_{1s}$ is a concave distribution. Since $F_{0s}$ is uniform, $F_s$ is also a concave distribution, and $F_s^{-1}$ is convex. Then the corollary follows.

## 5 Distributed Detection Algorithm

In this section we present the distributed detection algorithm. Our algorithm has the property that the communication cost scales with the number of sensors that observe an object, and not the total number of sensors in the SNET. We also present the equivalence of the distributed detection algorithm to its centralized counterpart, the BH procedure, by resorting to the switching relation [1] and Chernoff bound. First, we describe our distributed algorithm.

Observe that the BH procedure requires ordering of test statistics. Since ordering is not cost efficient in terms of communications, we use a sequential method to accomplish the linearly increasing thresholding of BH procedure. For reasons discussed earlier we consider communication complexity to be the number of broadcast messages.

**Distributed BH Algorithm:** First, each sensor obtains test statistics $y_s$ through the proposed transformation. Every sensor carries an indicator variable $\xi_s(t)$, such that $\xi_s(t) = 1$ if sensor $s$ has not transmitted a decision before iteration $t$, and $\xi_s(t) = 0$ otherwise. Sensors also carry a decision variable $\rho_s(t)$ such that $\rho_s(t) = 1$ if sensor $s$ decides $H_1$, and $\rho_s(t) = 0$ otherwise. At iteration $t$ each sensor has a threshold variable $l(i_t) = i_t \gamma / m$ and a bit counter $count_t$. Initialize $i_1 = 1$ and $count_0 = 0$. Then:

1. Sensor $s$ decides $H^s = H_1$ if $y_s \leq l(i_t)$ and $H^s = H_0$ otherwise. ($\rho_s(t)$ takes its corresponding value)
2. $s$ announces its decision to the network only if $\xi_s(t)\rho_s(t) = 1$
3. Assume $r_t$ sensors decide $H_1$ and declare to the network. Set $i_{t+1} = i_t + 1$ & $count_t = count_{t-1} + r_t$
4. If $count_t \geq i_t$ set variable $t_{max} = t$
5. If $i_t = m$ or $r_t = 0$ label sensors that declare $H^s = H_1$ until iteration $t_{max}$ as observing an object and quit algorithm, else go to step 1

The distributed algorithm described above leads to the same decision rule as the centralized BH procedure. However when there is a communication constraint of $\mathcal{C}$ messages, we only need to put a cap on the *count* variable and perform the distributed BH algorithm while $count_t \leq \mathcal{C}$. Observe that due to the concavity of $F_{1s}$ (Theorem 4.4), capping the *count* variable does not increase the $FDR$. In addition we argue that since the expected test statistics are samples of a convex function(Corollary 4.7), capping the *count* variable amounts to performing distributed detection algorithm with a smaller threshold $\gamma' \leq \gamma$, and hence the $FDR$ is generally smaller. Capping the count variable has an adverse effect in terms of fewer detections. Our main point here is to show that FDR and detection power gracefully degrades due to the monotonicity properties of the transformation. In other words smaller count does result in smaller detections but in a proportionate manner.

Figure 6(a) demonstrates this effect with a simple simulation study based on Monte Carlo simulations. For this demonstration we used $m_1 = 300$, $m = 1000$, $g_{0s} = N(0, 1)$, $g_{1s} = N(3, 1)$, with the transformation of Equation 2. We then varied $\mathcal{C}$, the communication bit budget, between 20 and 920 with increments of 100, and plotted the actual

FDR as a function of $\mathcal{C}$ for various values of FDR thresholds, $\gamma$. Figure 6(a) exhibits the FDR results of this empirical study. We will extensively simulate error rates (false alarms and misses) in Section 7. We note here that the detection power decreases accordingly when we cap the *count* variable. It was shown in [4] that the BH procedure controls FDR

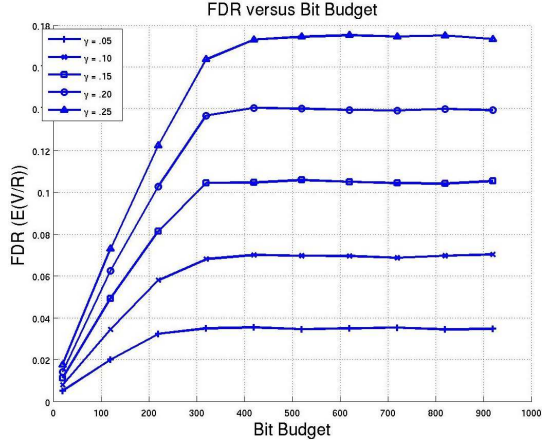
(a) Effect of capping the communications

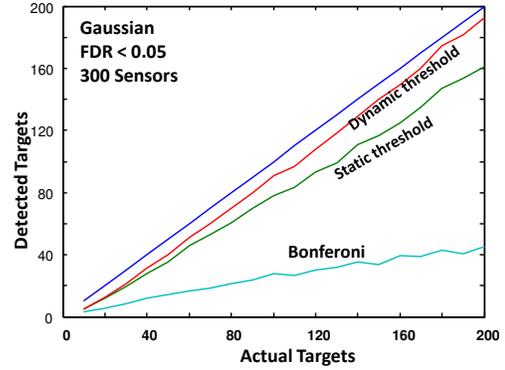
(b) Effect of Learning Object Density

Figure 6: Left: We capped the communication $count_t$ variable at $\{20, 120, 220, \ldots, 920\}$ for various FDR levels $\gamma$. Observe that the smaller the bit budget ($\mathcal{C}$), the smaller the actual FDR. Notice that FDR tapers off at levels $(m_0/m)\gamma$. Here $m_0 = 700$, $m = 1000$. Right: Threshold update through learning: The sequential estimate of $m_0$ allows for sequential update of threshold $\gamma$ for better detection rates. The distribution under null and significant hypotheses were $N(0,1)$ and $N(0,3)$ respectively.

at $\gamma m_0/m$. Depending on the unknown variable $m_0$, there may be an inherent conservatism in this strategy. We have analyzed an alternative strategy in [10], wherein at each update of the $count_t$ variable an estimate of actual number of objects, $\hat{m}_1$, is sequentially estimated based on the number of sensors that choose $H^s = H_1$ at stage, $t$. The threshold $l(i_t)$ is then adjusted based on the estimated object density. Our simulation results indicated that this strategy leads to a much better detection power, as seen in Figure 6(b).

### 5.1 Distributed BH Algorithm: Optimal Communication Cost with Centralized Performance

We next establish an important scaling property of the BH procedure. It is due to this property that we can limit the communication budget with an upper bound that depends on the number of sensors that have an object within their sensing range, and not the total number of sensors in the SNET.

**Theorem 5.1** *Let $m_1 = m - m_0$ be the number of sensors with significant hypothesis. The expected ratio of $m_1$ to the number of sensors that are declared to be significant (R) is lower bounded by $1 - \gamma$; i.e. $E\{m_1/R\} \geq 1 - \gamma$.*

Proof: We know that the BH procedure guarantees $E\{V/R\} = E\{V/(V+Z)\} \leq \gamma$. Evidently $Z \leq m_1$, meaning the number of correct detections cannot exceed the number of sensors with significant hypotheses. Then:

$$E\{V/(V+Z)\} = 1 - E\{Z/(V+Z)\} \Rightarrow E\{Z/(V+Z)\} \geq 1 - \gamma$$

$$1 - \gamma \leq E\{Z/(V+Z)\} \leq E\{m_1/(V+Z)\} = E\{m_1/R\}$$

which concludes the proof of this theorem. ∎

We caution the reader that the above bound does not guarantee detection performance. It only points to the fact that the number of eventual detections are generally smaller than the total number of actual objects.

**Equivalence to Centralized Algorithm:** The communication efficiency of our distributed BH procedure is that it is a first-crossing procedure, in contrast to the optimal centralized BH procedure, which is a last crossing procedure. Last crossing procedures are inherently inefficient. Although the number of eventual detections is typically smaller than the number of objects according to Theorem 5.1, we need to aggregate data from all the sensors in general to determine last crossing. On the other hand, by definition, the first crossing procedures must stop before last crossing. Moreover, the FDR guarantee on the first crossing is bounded from above by the last crossing. Consequently, the number of detections (and hence the communication efficiency) in first crossing procedure must be smaller than the actual number of objects. Nevertheless, there may be loss in performance. Below we argue that with our proposed transformation first and last crossing procedures lead to same performance. This establishes optimality of the distributed BH procedure in terms of both communication efficiency and performance.

In the asymptotic case the first-crossing and the last-crossing procedures are the same, and they terminate at the same decision point. This is because according to Corollary 4.7 the ordered test statistics are asymptotically samples of a convex function. However, this is not true in finite sample cases and therefore the first-crossing and the last-crossing procedures can have different termination points. Below we show that the distributed BH procedure achieves the last-crossing performance with high probability.

**Theorem 5.2** *Let $Y_{(k)}$ be the $k^{th}$ smallest test statistic. If $E(Y_{(\lceil \frac{k}{1-\epsilon} \rceil)}) \leq l_k$, then $Pr\{Y_{(k)} > l_k\}$ decays exponentially fast with $k$.*

Proof: See appendix. ∎

The implication of this theorem is that after a certain number of test statistics, say $k$, are tested against their corresponding thresholds, one can decide whether or not to continue the distributed algorithm with an exponentially small probability of error.

This result further suggests presetting $k$ tests at the beginning of the algorithm, which must be performed regardless of the outcome. Note, however, that $k$ can be fixed a priori and does not depend on the size of the SNET. We next show that such a modification does not affect important properties of our distributed algorithm.

**Theorem 5.3** *Consider the distributed detection algorithm with $k$ preset tests for suitably large $k$. Then:*
*(a) $FDR \leq \gamma$, and (b) With the expected number of messages equal to $\max\{k, E(\frac{m_1}{1-\gamma})\}$ the distributed algorithm achieves detection power of the centralized BH procedure with high probability.*

Proof: **a)** We show this part by showing that the distributed algorithm is in fact equivalent to the centralized algorithm, and that presetting $k$ tests affects only the communication cost. If there exists a $y_i \leq i\gamma/m$, $i \geq k$, then $k$ is immaterial. The upper bound then follows from Theorem 5.1. This is because the centralized algorithm would also map all test statistics less than $y_i$ to the significant hypothesis. Therefore $FDR \leq \gamma$ in this case. If there is no such $y_i$, $i \geq k$, then the algorithm chooses the largest test statistic $y_j \leq j\gamma/m$, $j < k$, and maps all smaller test statistics to the significant hypothesis. But here from Theorem 5.2 it follows that there is no other $j > k$ such that $y_i \leq i\gamma/m$, $i \geq k$ with high probability. Hence in both cases detection power of the centralized algorithm is achieved with high probability. ∎

## 6 Robustness Properties and Non-Ideal Sensing Model

Our development so far offers a solution to the distributed detection problem in the ideal model. Observe that since the families $\mathcal{G}_{0s}$ and $\mathcal{G}_{1s}$ are singletons, we can use $G_{0s}$ and $G_{1s}$ to define the transformation $Y_s = \chi_s(\mathbf{X}_s)$, and use the distributed BH procedure to perform detection. We have shown that this leads to optimal detection rate under the FDR criterion using BH procedure, and it has important scaling properties. In the non-ideal sensing model the sensors that are outside the radius $d_{min}$ receive a small residual signal from the objects. Since the received signal is not known, the exact distribution of observations are not available, i.e., $\mathcal{G}_{0s}$ and $\mathcal{G}_{1s}$ are no longer singletons. This leads to a deviation of $F_{0s}$ from $U[0,1]$, and $F_{0s}$ becomes a member of a family of distributions $\mathcal{F}_{0s}$. Similarly, $F_{1s}$ becomes a member of family of distributions, $\mathcal{F}_{1s}$.

In this section, we establish certain robustness properties of the proposed test statistics. These properties show that if we know $\mathcal{G}_{0s}$ and $\mathcal{G}_{1s}$ to within $\epsilon$ in terms of a certain distance measure, then we can identify the families $\mathcal{F}_{0s}$ and $\mathcal{F}_{1s}$ to within $\epsilon$ as well. We also establish that FDR scales gracefully when the distribution of $Y_{0s}$ deviates from $U[0,1]$ by $\epsilon$ under a suitable metric. Combining these results leads to efficient distributed detection for non-ideal sensing model with guarantees on performance. We begin by the robustness properties of the proposed test statistics.

**Theorem 6.1** *Let $\mu_{0s}$, $\hat{\mu}_{0s}$, and $\mu_{1s}$ be three measures such that $\mu_{1s} << \mu_{0s}$. Let $\mathbf{X}_s \sim \mu_{0s}$, $Y_s = \chi_s(\mathbf{X}_s) = \mu_{0s}(\{\mathbf{x} : \phi(\mathbf{x}) > \phi(\mathbf{X}_s)\})$, and $\hat{\mathbf{X}}_s \sim \hat{\mu}_{0s}$, $\hat{Y}_s = \chi_s(\hat{\mathbf{X}}_s) = \mu_{0s}(\{\mathbf{x} : \phi(\mathbf{x}) > \phi(\hat{\mathbf{X}})\})$. If $\sup_A | \mu_{0s}(A) - \hat{\mu}_{0s}(A) | \leq \epsilon \mu_{0s}(A)$, then $| \hat{F}_s(y_s) - y_s | \leq \epsilon y_s$ where $\hat{F}_s$ is the distribution of $\hat{Y}_s = \chi_s(\hat{\mathbf{X}}_s)$.*

Proof: See appendix. ∎

Note that we can obtain the robustness properties under significant hypothesis as well. To formalize that result we state the theorem here and omit its proof as it is a repetition of that of Theorem 6.1 with minor modifications.

**Theorem 6.2** *Let $\mu_{0s}$, $\hat{\mu}_{1s}$, and $\mu_{1s}$ be three measures such that $\mu_{1s} << \mu_{0s}$. Let $\mathbf{X}_s \sim \mu_{1s}$, $Y_s = \chi_s(\mathbf{X}_s) = \mu_{0s}(\{\mathbf{x} : \phi(\mathbf{x}) > \phi(\mathbf{X}_s)\})$, and $\hat{\mathbf{X}}_s \sim \hat{\mu}_{1s}$, $\hat{Y}_s = \chi_s(\hat{\mathbf{X}}_s) = \mu_{0s}(\{\mathbf{x} : \phi(\mathbf{x}) > \phi(\hat{\mathbf{X}})\})$. If $\sup_A | \mu_{1s}(A) - \hat{\mu}_{1s}(A) | \leq \epsilon \mu_{1s}(A)$, then $| \hat{F}_s(y_s) - F_s(y_s) | \leq \epsilon F_s(y_s)$ where $\hat{F}_s$ is the distribution of $\hat{Y}_s = \chi_s(\hat{\mathbf{X}}_s)$ and $F_s$ is the distribution of $Y_s = \chi_s(\mathbf{X}_s)$.*

Proof: Follows similarly to proof of Theorem 6.1. ∎

With these robustness properties of the test statistics, for continuous families $\mathcal{F}_{0s}$ such that $\mathcal{F}_{0s} = \{F_{0s} : |F_{0s}(y) - y| \leq \epsilon y\}$, we have an immediate non-asymptotic robustness result, which states that the FDR scales gracefully as a function of $\epsilon$.

**Theorem 6.3** *Let $Y_{0s}$ have continuous distribution $F_{0s}(y)$. If $|F_{0s}(y) - y| \leq \epsilon y$, the BH procedure bounds the false discovery rate by $\gamma(1 + \epsilon)$, i.e. $FDR \leq \gamma(1 + \epsilon)$.*

Proof: See appendix. ∎

The robustness result stated in Theorem 6.3 presents us with an immediate modification to the distributed BH algorithm in order to control FDR. It suggests that if we wish to control FDR at level $\gamma$, we only need to input the threshold $\gamma' = \gamma/(1 + \epsilon)$. The distributed BH algorithm with this modification can account for the non-ideal sensing model. Next using Theorem 6.2 we can establish a theorem parallel to Theorem 5.3. We omit the proof since it follows along the same lines as Theorem 5.3. The only modification is that each of the probability expressions are perturbed by a small amount on account of the perturbation of the underlying distributions.

**Theorem 6.4** *Consider the distributed detection algorithm with $k$ preset tests for suitably large $k$. Let the distributions satisfy the hypothesis of Theorem 6.1, 6.2. Then: (a) $FDR \leq \gamma'$, and (b) with expected number of messages equal to $\max\{k, E(\frac{m_1}{1-\gamma'})\}$ the distributed algorithm achieves detection power of the centralized BH procedure with high probability.*

# 7 Simulations

In this section we present some empirical studies based on Monte Carlo simulations on the proposed method. To obtain each data point in our simulation study we performed over 5000 monte-carlo iterations and found that this was sufficient to ensure confidence in our estimates. In this section we first show that the test statistics we propose performs better than other multi-dimensional transformations that have been proposed: namely, the radial transformation, the multidimensional counterpart of $p$ values, as well as the method proposed in [7]. Based on this result, we choose to use the proposed test statistics, and show that the BH procedure achieves a near Bayes Oracle error rate whereas Bonferroni procedure and Uncorrected testing cannot. We then present an SNET simulation, where we vary several parameters and examine the error rate and communication cost. For both studies we discuss the relevant parameters and setup in the corresponding sections.

## 7.1 Synthetic Data with Known Distributions

In our first study, we use the BH procedure to perform detection, and compare the error rate with that of Bayes Oracle. The setup is as follows: There are $m = 1000$ hypotheses tests, the number of objects $m_1$ is varied from $10\%$ to $90\%$. Under $H^s = H_0$, $\mathbf{X}_s \sim N(\mathbf{0}, I_3)$ and under $H^s = H_1$, $\mathbf{X}_s \sim N(\mathbf{1.5}, I_3)$. Here $I_3$ denotes the $3 \times 3$ identity matrix, $\mathbf{0}$ denotes the $3 \times 1$ zero vector, and $\mathbf{1.5}$ denotes the $3 \times 1$ vector $(1.5\ 1.5\ 1.5)'$, where $'$ denotes transpose operator. The test statistics are computed in three ways: first, we use the method proposed in this work, i.e., $Y_s = \mu_{0s}\{\mathbf{x} : \phi_s(\mathbf{x}) > \phi_s(\mathbf{X}_s)\}$, next we use the radial transformation, i.e., $Y_s = \mu_{0s}\{\mathbf{x} : g_{0s}(\mathbf{x}) \leq g_{0s}(\mathbf{X}_s)\}$, and finally we use dimension-by-dimension transformation proposed in [7]. In the dimension-by-dimension approach, the test statistics are calculated for each dimension separately by using the marginal distribution.

We next input these test statistics to the BH procedure to identify the tests where $H^s = H_1$. The FDR constraint is chosen to be $\gamma = .1$. Figure 7(a) presents the error rates associated with each of these methods versus $m_1/m$. Observe that the BH procedure with proposed test statistics comes close to the Bayes Oracle performance for high sparsity levels (small $m_1/m$). Furthermore, even at low sparsity levels it achieves the smallest error rate.

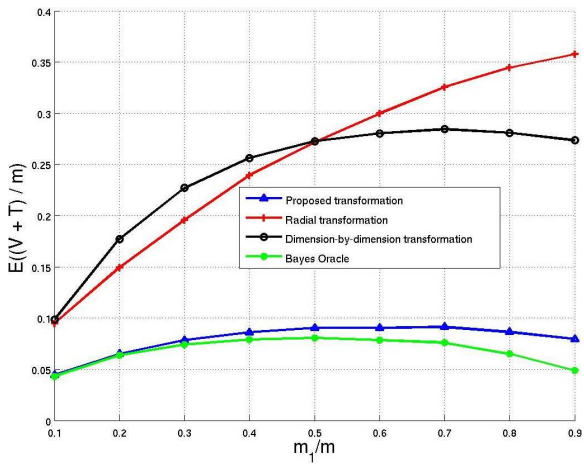 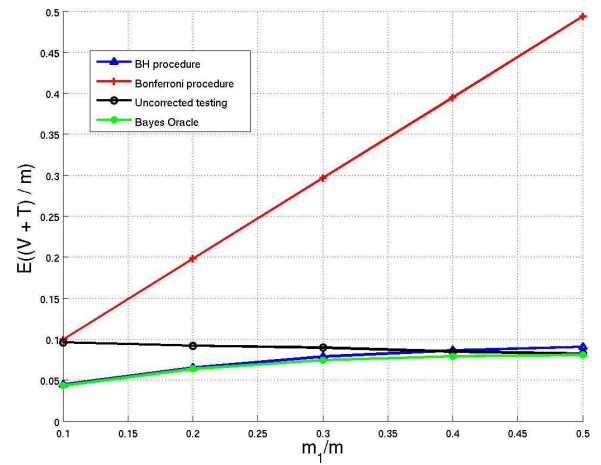

(a) Different Multi-dimensional Transformations    (b) Different Testing Procedures

Figure 7: Monte Carlo simulations for comparing Error rate vs object density for different multi-dimensional transformations and different decision criteria. BH procedure on the proposed test statistics dominates other transformations and testing strategies.

With the same setup, we now use the proposed test statistics and assess the error rate of the following detection schemes: BH procedure, Bonferroni procedure, and Uncorrected testing. The Bonferroni procedure takes the test statistics as input and tests if $Y_s \leq \gamma/m$ in order to decide which ones have $H^s = H_1$. Similarly, uncorrected testing checks if $Y_s \leq \gamma$. We compare the error rate of these three methods with that of the Bayes Oracle for $m_1$ varying from $10\%$ to $90\%$. The results clearly indicate that the performance of the BH procedure is close to that of the Bayes oracle. Figure 7(b) demonstrates these results. Note that the performance of BH procedure is strikingly similar to that of Bayes Oracle at high sparsity levels (low $m_1/m$).

## 7.2 SNET Simulation with Nonideal Model

In our next study we setup a SNET simulation to study the effects of SNET size, object density, and attenuation on the performance of the proposed method. We present the performance of the Bayes Oracle as a reference point when appropriate.

First we set up a $n \times n$ grid of sensors, where the distance between each sensor is 4 units. Then we place a number of objects at randomly chosen locations over the SNET, where the possible locations are at the center of grid squares. Figure 8 depicts this setup with a $3 \times 3$ grid and one object.

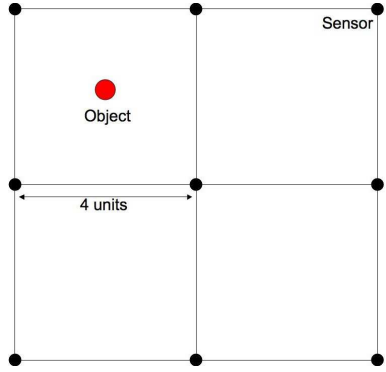

Figure 8: A $3 \times 3$ sensor grid and an object.

The observation model is given by

$$\mathbf{X}_s = \sum_t \frac{\boldsymbol{\theta}_t}{(d(s,t)/d_{min})^\alpha + 1} + \boldsymbol{\nu}_s$$

where $d(s,t)$ is the distance between sensor $s$ and an object $t$, $d_0 = 2\sqrt{2}$ is the distance between an object and the nearest sensor, $\alpha$ is the attenuation coefficient, and $\boldsymbol{\nu}_s \sim N(\mathbf{0}, I_3)$ is the noise. Here we have

$$\begin{aligned} H^s &= H_0: \ d(s,t) > d_0 \ \text{for all objects } t \\ H^s &= H_1: \ d(s,t) \leq d_0 \ \text{for an object } t \end{aligned}$$

We choose $d_{\min} = d_0$ for the attenuation model. This choice fixes the nominal signal-to-noise (SNR) ratio at the sensors. In other words in the presence of a single object in the entire sensor field, the sensor in the immediate vicinity of the object receives a signal with the same SNR irrespective of attenuation coefficient, $\alpha$. On the other hand objects not in the immediate vicinity suffer from path losses allowing us to study the impact of perturbations. Other choices for $d_{\min}$ are possible, however, they lead to scaling of both interference as well as nominal signal.

In our setup note that the second smallest distance between a sensor and an object is $\sqrt{6^2 + 2^2} = \sqrt{40}$ units. This implies that we have two candidates for nominal null distributions: (a) Nominal null distribution, $g_{0s}$ is a normal distribution with zero mean and noise variance; (b) Nominal null distribution $g_{0s}$ is a normal distribution with mean equal to signal received from a hypothetical object located at $\sqrt{40}$ units. For the significant hypothesis, we always assume the nominal distribution, $g_{1s} = N(\boldsymbol{\theta}_t, I_3)$. In our experiments we notice that error rates for two different nominal null distributions to be similar. The differences appeared to be in the composition of false alarms and misses. This is because the second assumption is conservative, i.e., a distant object is assumed even if there does not exist any object. Our simulations allowed more than one object in the immediate vicinity of a sensor. However, we did not notice any degradation in performance.

**Effect of Object Density:** In our first study we have $n = 25$, which leads to $m = 625$ sensors. We choose $\gamma = 0.1$, $\boldsymbol{\theta}_t = (2\ 2\ 2)'$, and $\alpha = 2$. We then vary the number of objects such that $m_1/m \in \{.03, .06, \ldots, .15\}$, and observe the error rate and communication cost of distributed BH procedure, Bonferroni procedure, Uncorrected testing, and Bayes Oracle. Figures 9(a) and 9(b) demonstrate the results of this study. Notice that the error rate of the distributed BH procedure again closely tracks that of the Bayes Oracle. However, while the Bayes Oracle uses the knowledge of $m_1/m$, and the actual distribution of observations at each sensor, the distributed BH procedure only uses the assumed distributions.

The expected proportion of communication cost to $m_1$ remains near 1 for the distributed BH procedure, whereas it significantly deviates from 1 for Bonferroni procedure and Uncorrected testing. This is because, the Bonferroni procedure, due to its stringent threshold $\gamma = 0.1/625$, misses most of the sensors that are within $d_{min}$ of an object. On the other extreme, the Uncorrected testing suffers from a large number of false alarms, which is a constant proportion of $m$, and therefore the communication cost is significantly larger than $m_1$.

**Effect of Attenuation Coefficient:** We next study how the attenuation coefficient $\alpha$ affects the error rates and communication costs of the competing schemes. For this setup we again have $n = 25$, which leads to $m = 625$ sen-sors, and we choose $\gamma = 0.1$ and $\boldsymbol{\theta}_t = (1.5\ 1.5\ 1.5)'$. The object density is fixed, where $m_1/m = 0.1$. We vary $\alpha \in \{2, 2.2, 2.4, \ldots, 4\}$ and observe the error rate and communication cost versus $\alpha$. Intuitively, as we increase the attenuation coefficient, the distributions $g_{0s} = N(\hat{\boldsymbol{\theta}}_t, I_3)$ and $g_{1s} = N(\boldsymbol{\theta}_t, I_3)$ become more separable. This in turn is expected to decrease the error rate, and increase the detection rate. Increasing the detection rate increases the communication costs. These are precisely the effects we observe in Figures 10(a) and 10(b).

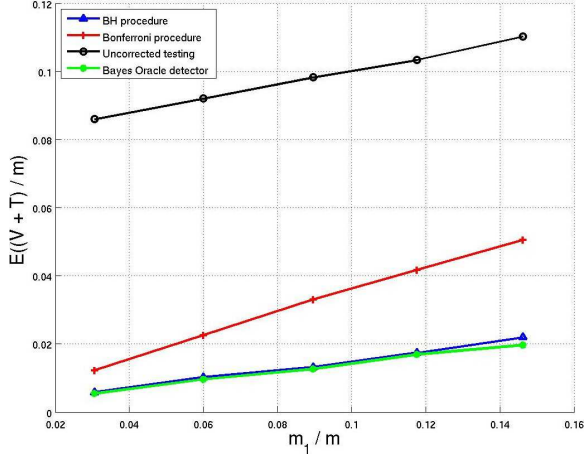
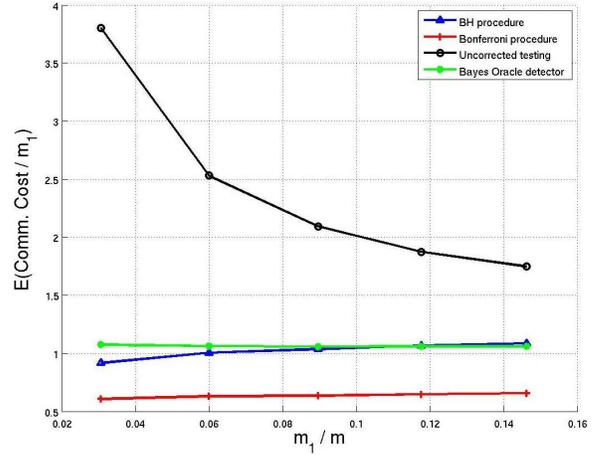

(a) Error vs. Target Density

(b) Comm. Cost vs. Target Density

Figure 9: Monte Carlo simulations for comparing error rate vs object density for the proposed statistics in the SNET setup: the distributed BH procedure using the proposed test statistics achieves the minimal error rate and closely tracks the performance of Bayes Oracle. Here $m = 625$, $\boldsymbol{\theta}_t = (2\ 2\ 2)'$, $\boldsymbol{\nu}_s \sim N(\mathbf{0}, I_3)$, $\alpha = 2$.

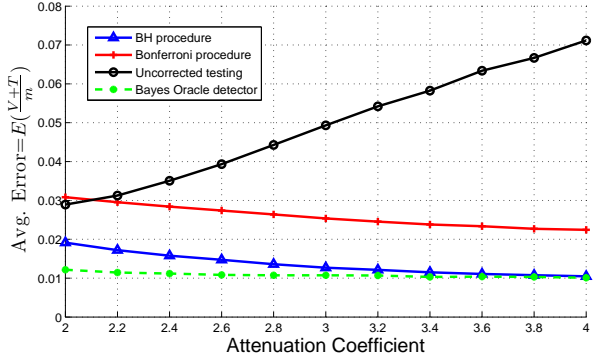
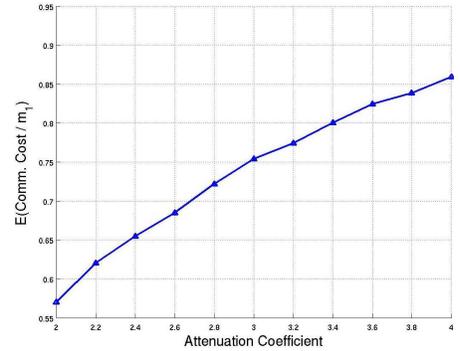

(a) Error Rate vs. Attenuation

(b) Comm. Cost vs. Attenuation

Figure 10: Monte Carlo comparison of error rate & communication cost as a function of attenuation coefficient ($\alpha$) for different strategies. Monte-Carlo simulations with 625 multi-modal sensors, with 62 sensors in immediate vicinity of a target were simulated. The parameters governing the sensing model were $\boldsymbol{\theta}_t = (1.5\ 1.5\ 1.5)'$, with $\boldsymbol{\nu}_s \sim N(\mathbf{0}, I_3)$. The proposed distributed BH procedure achieves the minimal error rate and closely tracks the performance of Bayes Oracle when we use the proposed test statistics. As $\alpha$ increases, the distributions become more separable and the error rate decreases. Note that as $\alpha$ increases, the distributions become more separable, which in turn increases the detection rate and associated communication cost.

**Effect of SNET size:** In our final study we examine the size of the SNET on the communication costs. What we wish to do is to fix the number of objects and increase the size of the SNET grid. The effect we wish to show is that for the BH procedure, no matter the size of the SNET, the communication cost scales with $m_1$, the number of sensors that are in the vicinity of an object, and not $m$, the size of the SNET. For this study we fix $\alpha = 2$, $\gamma = 0.1$, $m_1 = 60$, and $\boldsymbol{\theta}_t = (2\ 2\ 2)'$. We then vary $n \in \{25, 35, 45, 55\}$, which leads to $m \in \{625, 1225, 2025, 3025\}$. Figure 11 demonstrates the results of this study. Observe that for the uncorrected testing the communication cost linearly increases as a function of the SNET size, whereas the distributed BH procedure is able to retain a near constant fraction of communication cost to $m_1$. Notice that Bonferroni procedure has the lowest communication cost, however this is due to the fact that detection rate of Bonferroni procedure is very small.

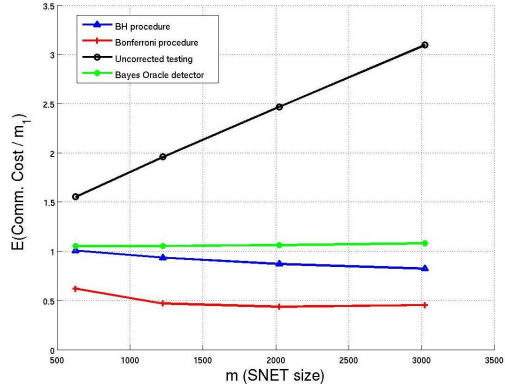

Figure 11: Monte Carlo simulation of communication cost/$m_1$ with SNET size ($m$) using the proposed statistics for the SNET setup: as $m$ increases, the communication cost of the Uncorrected testing increases whereas the distributed BH procedure retains a near constant fraction of communication costs to $m_1$. Bonferroni procedure has the lowest communication cost, however this is due to the detection rate being very small. Here $\alpha = 2$, $\gamma = 0.1$, $m_1 = 60$, and $\boldsymbol{\theta}_t = (2\ 2\ 2)'$.

## 8 Conclusion

In this paper we developed tools for detection of localized events, sources, or abnormalities within SNETs. Unlike decentralized detection where the information is globally available, the focus here was on problems, where only a small number of sensors in the vicinity of the phenomena are in the field of observation. We call these problems local information problems. For such problems the main difficulty arises from the coupling of: **a)** uncertainty in the number of events, sources or abnormalities and uncertainty in the possible locations; **b)** multiplicity of false alarms. Although not evident at first sight, these fundamental difficulties call for collaboration in the SNET in order to meet global constraints.

We proposed FDR as a performance criterion for local information problems in SNETs. The reasoning behind FDR was the fact that FDR adapts to the unknown object density, which is of great importance for distributed detection problems. Namely, we do not know not only how many events take place at any time, but also where these events occur. The adaptive nature of FDR made it a very valuable tool to address these issues.

We next introduced a transformation that maps multidimensional observations to single dimensional test statistics, which has important properties for distributed algorithms. Namely, asymptotically the ordered test statistics are samples of a convex function. This allowed us to devise a distributed BH procedure, which is a first crossing procedure that also has desirable scaling properties in terms of the communication costs. Namely, the communication cost of the distributed algorithm scales with the number of significant sensors (sensors in the close vicinity of an object), and not the whole SNET. We also showed that the distributed BH procedure achieves the performance of its centralized counterpart.

We quantified robustness of the distributed algorithm and the proposed transformation to unknown perturbations in the nominal distribution. This issue is particularly relevant in a sensing field where the path losses and attenuation coefficients are not known. The simulation studies confirmed this assertion by demonstrating that distributed BH procedure tracks the performance of the Bayes Oracle in terms of the error rate, even in the non-ideal model, with communication costs scaling with the number of significant sensors.

# Appendix

## Proof of Theorem 2.1

First note that from Lagrangian duality it follows that,

$$\gamma_w = \min_u \max_{H^s} \left( \Pr\{V \geq 1 \mid \{H^s : s \in \mathcal{S}\}\} + \Pr\{T \geq 1 \mid \{H^s : s \in \mathcal{S}\}\} \right) \geq \max_{\Pr\{(H^s): s \in \mathcal{S}\}} \min_u \Pr(V \geq 1) + \Pr(T \geq 1)$$

where, we can substitute any prior distribution for $\Pr\{(H^s) : s \in \mathcal{S}\}$. Consequently, we are left to establish a bound for the Bayesian problem. Now we observe that the error event,

$$\mathcal{E} = \{u(X_1, X_2, \ldots, X_m) \neq \{H^s : s \in \mathcal{S}\}\} = \{V \geq 1\} \cup \{T \geq 1\}$$

Therefore, from Fano's inequality it follows that for any strategy $u(\cdot)$:

$$\Pr(V \geq 1) + \Pr(T \geq 1) \geq \Pr(\mathcal{E}) \geq \frac{1}{m}\Phi\{H^s : s \in \mathcal{S}\} \mid X_1, X_2, \ldots, X_m) - \frac{1}{m} = \Phi(H^s \mid X_s) - \frac{1}{m}$$

where $\Phi\{H^s : s \in \mathcal{S}\}$ is the conditional entropy. The last equality follows by substituting a independent Bernoulli prior for presence or absence of objects. ∎

## Proof of Theorem 4.1

Note that for any sequence $\phi_1 > \phi_2 > \ldots$ the sets $A_i = \{\mathbf{x} : \phi_s(\mathbf{x}) > \phi_i\}$ form a nested sequence of sets such that $A_1 \subset A_2 \subset \ldots$. Then

$$\begin{aligned}
\Pr\{Y_s \leq y_s\} &= \Pr\{\mu_{0s}\{\mathbf{x} : \phi_s(\mathbf{x}) > \phi_s(\mathbf{X}_s)\} \leq \mu_{0s}\{\mathbf{x} : \phi_s(\mathbf{x}) > \phi_s(\mathbf{x}_s)\}\} \\
&= \Pr\{\mu_{0s}\{\mathbf{x} : \phi_s(\mathbf{x}) > \phi_s(\mathbf{X}_s)\} < \mu_{0s}\{\mathbf{x} : \phi_s(\mathbf{x}) > \phi_s(\mathbf{x}_s)\}\} \\
&= \Pr\{\{\mathbf{x} : \phi_s(\mathbf{x}) > \phi_s(\mathbf{X}_s)\} \subset \{\mathbf{x} : \phi_s(\mathbf{x}) > \phi_s(\mathbf{x}_s)\}\} \\
&= \Pr\{\phi_s(\mathbf{X}_s) > \phi_s(\mathbf{x}_s)\} = \Pr\{\mathbf{x} : \phi_s(\mathbf{x}) > \phi_s(\mathbf{x}_s)\} = \mu_{0s}\{\mathbf{x} : \phi_s(\mathbf{x}) > \phi_s(\mathbf{x}_s)\} = y_s
\end{aligned}$$

where the probability measure is $\mu_{0s}$, the second inequality follows from the continuity of $Y_s$, and the third equality follows from the fact that the sets are nested. The independence of the test statistics under null hypothesis follows from our conditional independence assumptions of Section 3. ∎

## Proof of Theorem 4.3

We can write for $Y_s$ and $Z_s$:

$$\begin{aligned}
\Pr\{Y_s \leq \gamma_s\} &= \pi\Pr\{Y_s \leq \gamma_s \mid H^s = H_1\} + (1-\pi)\Pr\{Y_s \leq \gamma_s \mid H^s = H_0\} \\
&= \pi\Pr\{Y_s \leq \gamma_s \mid H^s = H_1\} + (1-\pi)\gamma_s \\
\Pr\{Z_s \leq \gamma_s\} &= \pi\Pr\{Z_s \leq \gamma_s \mid H^s = H_1\} + (1-\pi)\Pr\{Z_s \leq \gamma_s \mid H^s = H_0\} \\
&= \pi\Pr\{Z_s \leq \gamma_s \mid H^s = H_1\} + (1-\pi)\gamma_s
\end{aligned}$$

Then, to prove our result, it suffices to show that $\Pr\{Y_s \leq \gamma_s \mid H^s = H_1\} \geq \Pr\{Z_s \leq \gamma_s \mid H^s = H_1\}$. To show this, let $A^{\chi_s} = \{\mathbf{x} : \phi_s(\mathbf{x}) > \phi_1\}$ be the set such that $\mu_{0s}A^{\chi_s} = \gamma_s$. Notice that for $Z_s$, the uniform distribution under the null hypothesis assumption implies $\Pr\{z : z \leq \gamma_s \mid H^s = H_0\} = \mu_{0s}\{\mathbf{x} : \hat{\chi}_s(\mathbf{x}) \leq \gamma_s \mid H^s = H_0\} = \gamma_s$. Write $A^{\hat{\chi}_s} = \{\mathbf{x} : \hat{\chi}_s(\mathbf{x}) \leq \gamma_s \mid H^s = H_0\}$. Then, $\mu_{1s}A^{\chi_s} = \mu_{1s}(A^{\chi_s} - A^{\hat{\chi}_s}) + \mu_{1s}(A^{\chi_s} \cap A^{\hat{\chi}_s})$, and similarly $\mu_{1s}A^{\hat{\chi}_s} = \mu_{1s}(A^{\hat{\chi}_s} - A^{\chi_s}) + \mu_{1s}(A^{\chi_s} \cap A^{\hat{\chi}_s})$, where $A - B$ denotes the removal of set $B$ from set $A$.

Observe that showing $\Pr\{Y_s \leq \gamma_s \mid H^s = H_1\} \geq \Pr\{Z_s \leq \gamma_s \mid H^s = H_1\}$ is equivalent to showing $\mu_{1s}A^{\chi_s} - \mu_{1s}A^{\hat{\chi}_s} = \mu_{1s}(A^{\chi_s} - A^{\hat{\chi}_s}) - \mu_{1s}(A^{\hat{\chi}_s} - A^{\chi_s}) \geq 0$. To show this, observe that $\mu_{0s}(A^{\chi_s} - A^{\hat{\chi}_s}) = \mu_{0s}(A^{\hat{\chi}_s} - A^{\chi_s}) = \gamma'$ for some $\gamma' = \gamma - \mu_{0s}(A^{\chi_s} \cap A^{\hat{\chi}_s})$. But, over $A^{\chi_s} - A^{\hat{\chi}_s}$, $d\mu_{1s}/d\mu_{0s} = \phi_s > \phi_1$, and hence $\mu_{1s}(A^{\chi_s} - A^{\hat{\chi}_s}) > \phi_1\gamma'$. Similarly, over $A^{\hat{\chi}_s} - A^{\chi_s}$, $d\mu_{1s}/d\mu_{0s} = \phi_s \leq \phi_1$, and hence $\mu_{1s}(A^{\hat{\chi}_s} - A^{\chi_s}) \leq \phi_1\gamma'$, which implies $\mu_{1s}(A^{\chi_s}) - \mu_{1s}(A^{\hat{\chi}_s}) \geq 0$ and concludes the proof. ∎

## Proof of Theorem 4.4

Again noting that for any sequence $\phi_1 > \phi_2 > \ldots$ the sets $A_i = \{\mathbf{x} : \phi_s(\mathbf{x}) > \phi_i\}$ form a nested sequence of sets such that $A_1 \subset A_2 \subset \ldots$, we can write:

$$\begin{aligned} F_{1s}(y_s) &= \Pr\{Y_s \leq y_s\} \\ &= \Pr\{\mu_{0s}\{\mathbf{x} : \phi_s(\mathbf{x}) > \phi_s(\mathbf{X}_s)\} \leq \mu_{0s}\{\mathbf{x} : \phi_s(\mathbf{x}) > \phi_s(\mathbf{x}_s)\}\} \\ &= \Pr\{\mu_{0s}\{\mathbf{x} : \phi_s(\mathbf{x}) > \phi_s(\mathbf{X}_s)\} < \mu_{0s}\{\mathbf{x} : \phi_s(\mathbf{x}) > \phi_s(\mathbf{x}_s)\}\} \\ &= \Pr\{\{\mathbf{x} : \phi_s(\mathbf{x}) > \phi_s(\mathbf{X}_s)\} \subset \{\mathbf{x} : \phi_s(\mathbf{x}) > \phi_s(\mathbf{x}_s)\}\} \\ &= \Pr\{\phi_s(\mathbf{X}_s) > \phi_s(\mathbf{x}_s)\} = \Pr\{\mathbf{x} : \phi_s(\mathbf{x}) > \phi_s(\mathbf{x}_s)\} = \mu_{1s}\{\mathbf{x} : \phi_s(\mathbf{x}) > \phi_s(\mathbf{x}_s)\} \end{aligned}$$

Observe that here the probability measure is $\mu_{1s}$, because $\mathbf{X}_s$ is sampled with respect to $G_{1s}$.

Now, let $\phi_1 > \phi_2 > \phi_3$ be such that for $A_i = \{\mathbf{x} : \phi_s(\mathbf{x}) > \phi_i\}$, $i = 1, 2, 3$, $\mu_{0s}(A_1) = y_s^1$, $\mu_{0s}(A_2) = y_s^2 = y_s^1 + \delta_0$, and $\mu_{0s}(A_3) = y_s^3 = y_s^2 + \delta_0$ for some appropriate $y_s^1$ and $\delta_0$. Also, $F_{1s}(y_s^1) = \mu_{1s}(A_1) = z_s^1$, $F_{1s}(y_s^2) = \mu_{1s}(A_2) = z_s^2 = z_s^1 + \delta_1$, and $F_{1s}(y_s^3) = \mu_{1s}(A_3) = z_s^3 = z_s^2 + \delta_2$ for some appropriate $z_s^1$, $\delta_1$ and $\delta_2$.

Notice that $\delta_1 = \mu_{1s}(A_2 - A_1)$ and $\delta_2 = \mu_{1s}(A_3 - A_2)$. Noting that $\mu_{0s}(A_2 - A_1) = \mu_{0s}(A_3 - A_2) = \delta_0$ and noting that $A_i$ are constructed using the Radon-Nikodym derivative, it follows that $\delta_1 \geq \delta_2$. Thus we can write

$$\frac{F_{1s}(y_s^2) - F_{1s}(y_s^1)}{y_s^2 - y_s^1} = \frac{\delta_1}{\delta_0} \geq \frac{\delta_2}{\delta_0} = \frac{F_{1s}(y_s^3) - F_{1s}(y_s^2)}{y_s^3 - y_s^2}$$

But this holds true for all $\phi_1 > \phi_2 > \phi_3$ such that $\delta_0 > 0$, and hence the result follows. ∎

## Proof of Theorem 5.2

Let $N_k = \#\{j : y_j \leq l_k\} = \sum_{j=1}^m I_{\{y_j \leq l_k\}}$. By the switching relation (see for example [1]) the following relationship holds for any $k$: $\{E(y_{(\lceil \frac{k}{1-\epsilon} \rceil)}) \leq l_k\} \Leftrightarrow \{E(N_k) \geq \lceil \frac{k}{1-\epsilon} \rceil\}$. Therefore, $E(y_{(\lceil \frac{k}{1-\epsilon} \rceil)}) \leq l_k \Rightarrow E(N_k) \geq \frac{k}{1-\epsilon}$ and $k \leq E(N_k)(1-\epsilon)$.

$$\begin{aligned} \Pr\{Y_{(k)} > l_k\} = \Pr\{N_k < k\} &\leq \Pr\{N_k < E(N_k)(1-\epsilon)\} \\ &\leq \exp\{-\frac{\epsilon^2 E(N_k)}{2}\} \quad (4) \\ &\leq \exp\{-\frac{\epsilon^2 k}{2(1-\epsilon)}\} \quad (5) \end{aligned}$$

Inequality 4 follows from the Chernoff bound, and inequality 5 follows from the application of switching relation along with the assumption of the theorem. ∎

## Proof of Theorem 6.1

We know from Theorem 4.1 that $Y_s$ is uniformly distributed in $[0, 1]$. Similarly to the development of that theorem,

$$\begin{aligned} \Pr\{\hat{Y}_s \leq y_s\} &= \Pr\{\mu_{0s}\{\mathbf{x} : \phi_s(\mathbf{x}) > \phi_s(\hat{\mathbf{X}}_s)\} \leq \mu_{0s}\{\mathbf{x} : \phi_s(\mathbf{x}) > \phi_s(\mathbf{x}_s)\}\} \\ &= \Pr\{\mu_{0s}\{\mathbf{x} : \phi_s(\mathbf{x}) > \phi_s(\hat{\mathbf{X}}_s)\} < \mu_{0s}\{\mathbf{x} : \phi_s(\mathbf{x}) > \phi_s(\mathbf{x}_s)\}\} \\ &= \Pr\{\{\mathbf{x} : \phi_s(\mathbf{x}) > \phi_s(\hat{\mathbf{X}}_s)\} \subset \{\mathbf{x} : \phi_s(\mathbf{x}) > \phi_s(\mathbf{x}_s)\}\} \\ &= \Pr\{\phi_s(\hat{\mathbf{X}}_s) > \phi_s(\mathbf{x}_s)\} = \Pr\{\mathbf{x} : \phi_s(\mathbf{x}) > \phi_s(\mathbf{x}_s)\} \end{aligned}$$

Here the probability measure is $\hat{\mu}_{0s}$, since $\hat{\mathbf{X}}_s$ is drawn with respect to that measure. Then, $\hat{F}_{0s}(y_s) = \Pr\{\hat{Y}_s \leq y_s\} = \hat{\mu}_{0s}\{\mathbf{x} : \phi_s(\mathbf{x}) > \phi_s(\mathbf{x}_s)\}$. However, by hypothesis of the theorem, $\sup_A | \mu_{0s}(A) - \hat{\mu}_{0s}(A) | \leq \epsilon \mu_{0s}(A)$, and hence $| \hat{\mu}_{0s}\{\mathbf{x} : \phi_s(\mathbf{x}) > \phi_s(\mathbf{x}_s)\} - \mu_{0s}\{\mathbf{x} : \phi_s(\mathbf{x}) > \phi_s(\mathbf{x}_s)\} | \leq \epsilon \mu_{0s}\{\mathbf{x} : \phi_s(\mathbf{x}) > \phi_s(\mathbf{x}_s)\}$. Finally noting that $\mu_{0s}\{\mathbf{x} : \phi_s(\mathbf{x}) > \phi_s(\mathbf{x}_s)\} = y_s$, we have the result. ∎

**Proof of Theorem 6.3**

Define $\gamma_k = k\gamma/m$. Let $Y_{0s}$, $s = 1, 2, \ldots, m_0$ be the $m_0$ test statistics under null hypothesis. Denote with $C_s(k)$ the event that if $Y_{0s}$ mapped to $H^s = H_1$, exactly $k-1$ other test statistics are mapped to $H_1$. Then;

$$
\begin{aligned}
E(V/R) &= \sum_{s=1:m_0} \sum_{k=1:m} \frac{1}{k} \Pr\{Y_{0s} \leq \gamma_k, C_s(k)\} = \sum_{s=1:m_0} \sum_{k=1:m} \frac{1}{k} \Pr\{Y_{0s} \leq \gamma_k\} \Pr\{C_s(k)\} \\
&\leq \sum_{s=1:m_0} \sum_{k=1:m} \frac{1}{k} \frac{\gamma k}{m} (1+\epsilon) \Pr\{C_s(k)\} = (1+\epsilon) \sum_{s=1:m_0} \frac{\gamma}{m} \sum_{k=1:m} \Pr\{C_s(k)\} \\
&= (1+\epsilon) \sum_{s=1:m_0} \frac{\gamma}{m} = (1+\epsilon) \frac{\gamma m_0}{m} \leq (1+\epsilon)\gamma
\end{aligned}
$$

The second equality follows because $Y_{0s}$ is independent of all other test statistics. ∎

# References

[1] F. Abramovich, Y. Benjamini, D. Donoho, and I. Johnstone. Adapting to unknown sparsity by controlling the false discovery rate. *Technical Report, Statistics Department, Stanford University*, 2000.

[2] S. Aeron, M. Zhao, and V. Saligrama. Algorithms and bounds for sensing capacity and compressed sensing with applications to learning graphical models. *Information Theory and Applications Workshop, 2008*, pages 303–309, 27 2008-Feb. 1 2008.

[3] S. Appadwedula, V. V. Veeravalli, and D. L. Jones. Robust and locally-optimum decentralized detection with censoring sensors. In *5th Int. Conf. on Information Fusion*, 2002.

[4] Y. Benjamini and Y. Hochberg. Controlling the false discovery rate: A practical and powerful approach to multiple testing. *Journal of the Royal Statistical Society, Series B*, 57, 1995.

[5] Y. Benjamini and D. Yekuteli. The control of the false discovery rate in multiple testing under dependency. *The Annals of Statistics*, 29, 2001.

[6] J.F. Chamberland and V. Veeravalli. Asymptotic results for decentralized detection in power constrained wireless sensor networks. *IEEE Special Issue on Wireless Sensor Networks*, pages 1007–1015, August 2004.

[7] Z. Chi. False discovery rate control with multivariate p-values. Technical Report TR 06-10, University of Connecticut Department of Statistics.

[8] P-J Chung, J. F. Böhme, C. F. Mecklenbrauker, and A. O. Hero. Detection of the number of signals using the benjamini-hochberg procedure. *IEEE Trans on Signal Processing*, 55(5):2497–2508, 2007.

[9] H. A. David and H. N. Nagaraja. *Order Statistics*. John Wiley and Sons, 2003.

[10] E. Ermis and V. Saligrama. Dynamic thresholding for distributed multiple hypotheses testing. In *IEEE Statistical Signal Processing Workshop*, Madison, WI, 2007.

[11] E. B. Ermis, M. Alanyali, and V. Saligrama. Search and discovery in an uncertain networked world. *IEEE Signal Processing Magazine*, 23:107–118, 2006.

[12] E. B. Ermis and V. Saligrama. Adaptive statistical sampling methods for decentralized estimation and detection of localized phenomena. In *IPSN '05: Proceedings of the 4th international symposium on Information processing in sensor networks*, page 19, Piscataway, NJ, USA, 2005. IEEE Press.

[13] E. B. Ermis and V. Saligrama. Adaptive statistical sampling methods for decentralized estimation and detection of localized phenomena. *Acoustics, Speech, and Signal Processing, 2005. Proceedings. (ICASSP '05). IEEE International Conference on*, 5:v/1045–v/1048 Vol. 5, March 2005.

[14] E. B. Ermis and V. Saligrama. Detection and localization in sensor networks using distributed fdr. *Information Sciences and Systems, 2006 40th Annual Conference on*, pages 699–704, March 2006.

[15] E. B. Ermis and V. Saligrama. Robust distributed detection with limited range sensors. *Acoustics, Speech and Signal Processing, 2007. ICASSP 2007. IEEE International Conference on*, 2:II–1009–II–1012, April 2007.

[16] C. Genovese and L Wasserman. Operating characteristics and extensions of the false discovery rate procedure. *Journal of the Royal Statistical Society, Series B*, 64:479–498, 2002.

[17] Gregory, CJ and Carthy, RR and Pearlstine, LG. Survey and monitoring of species at risk at camp blanding training site, northeastern florida. *Southeastern Naturalist*, pages 473–498, 2006.

[18] T. Kasetkasem and P. K. Varshney. Communication structure planning for multisensor detection systems. *IEE Proc. Radar, Sonar Navig.*, 48:2–8, 2001.

[19] E. L. Lehmann. *Testing Statistical Hypothesis*. John Wiley and Sons, 1986.

[20] M. Leoncini, G. Resta, and P. Santi. Analysis of a wireless sensor dropping problem in wide-area environmental monitoring. *Information Processing in Sensor Networks, 2005. IPSN 2005. Fourth International Symposium on*, pages 239–245, April 2005.

[21] M. Longo, T. D. Lookabaugh, and Robert M. Gray. Quantization for decentralized hypothesis testing under communication constraints. *IEEE Trans. on Inform. Theory*, 36:241–255, 1990.

[22] N. Patwari and A. Hero. Reducing transmissions from wireless sensors in distributed detection networks using hierarchical censoring. In *ICASSP*. IEEE, 2003.

[23] H. V. Poor and G. W. Wornell, editors. *Wireless Communication*. Prentice Hall PTR, Upper Saddle River, NJ, USA, 1998.

[24] C. Rago, P. Willett, and Y. Bar-Shalom. Censoring sensors: a low-communication-rate scheme for distributed detection. *IEEE Trans. Aerosp. Electron. Syst.*, 32:554–568, 1996.

[25] C. D. Scott and E. Kolaczyk. Annotated minimum volume sets for nonparametric anomaly discovery. *Statistical Signal Processing, 2007. SSP '07. IEEE/SP 14th Workshop on*, pages 234–238, Aug. 2007.

[26] P.F. Swaszek and P. Willett. Parley as an approach to distributed detection. *IEEE Transactions on Aerospace and Electronic Systems*, pages 447–457, January 1995.

[27] H. L. Van Trees. *Detection Estimation and Modulation Theory*. John Wiley and Sons, 1968.

[28] J. N. Tsitsiklis. Decentralized detection. *in Advances in Statistical Signal Processing, H. V. Poor and J. B. Thomas Eds*, 2, 1993.

[29] V. Saligrama, M. Alanyali, O. Savas. Distributed detection in sensor networks with packet losses and finite capacity links. *IEEE Transactions on Signal Processing*, November 2006.

[30] P. K. Varshney. *Distributed Detection and Data Fusion*. Springer, 1997.

[31] L. Xiao, S. Boyd, and S. Lall. A scheme for robust distributed sensor fusion based on average consensus. In *IPSN*, 2005.